\documentstyle[aps,12pt]{revtex}

\newcommand{\fett}[1]{\mbox{\mathversion{bold}$#1$\mathversion{normal}}}

\newcommand{\light}[1]{\tilde {\bf #1}}

\begin{document}  

\title{Nucleon form factors in a light-front framework:\\
a model with instanton-induced forces}

\author{ Luis Prado, Jr.\thanks{e-mail: prado@itkp.uni-bonn.de \protect\newline
\indent  Tel.: +49 (0)228 73 2372  \protect\newline
\indent Fax: +49 (0)228 73 3728}\\
{\sl Institut f\"ur Theoretische Kernphysik,\\ Universit\"at Bonn,
 Nu{\ss}allee 14--16, D-53115 Bonn, FRG}}

\date{\today}

\maketitle
\bigskip
\begin{center}
{\large Abstract}
\end{center}
\bigskip
\begin{abstract}
Nucleon form factors are evaluated in the spacelike region in a 
light-front framework. Our phenomenological constituent quark model is 
based on a relativistic formulation of the Hamiltonian dynamics. 
The baryon  dynamics is solved in the nucleon rest frame 
with relativistic kinetic energy, linearly rising 
confining potential and a residual interaction based on instanton-induced 
forces, the so-called 't Hooft interaction. 
The wave funtions of the model are used to compute the nucleon form 
factors in the impulse approximation. In spite of taking no
phenomenological constituent quark form factor by the one-body 
electromagnetic current, we obtain a very good description of all 
electromagnetic form factors for momentum transfers up to $-$3 GeV$^2$.
The effect of the 't Hooft interaction is carefully examined in the model.
\end{abstract}

$\;$

PACS number(s): 
12.39.Ki, 
13.40.Gp, 
12.40.Yx, 
11.10.St, 
14.20.Dh  
\newpage

\section{Introduction}
The investigation of electromagnetic form factors has doubtless 
a crucial importance to gain qualitative insight into the 
hadronic structure. Recently, the nucleon electromagnetic form factors 
have received  much theoretical attention due to new experimental results
 [1-3] and several proposed experiments, 
which shall be realized in near future [4]. 

Among the most recent relativistic constituent quark models, 
we should high light the great achievements of 
the semibosonized Nambu-Jona-Lasinio model [5-7], 
the quark-diquark instantaneous Bethe-Salpeter (BS) framework [8], 
the light-front framework with simple nucleon wave funtions [9,10]
and the light-front relativistic Hamiltonian dynamics (RHD) [11,12]. 

In the refered investigations within the RHD, the
eigenvalue equation for the mass operator has been identified with 
the Capstick-Isgur Hamiltonian [13], which contains an interaction 
term composed of a confining part and a one-gluon-exchange 
(OGE) part.  We will also use the light-front RHD [14], but our
principal aim throughout this paper is to examine the influence 
of instanton-induced forces on the nucleon form factors.  

Instanton effects have been computed by 't Hooft and others [15-17]
in the soft-QCD regime. Those effects yields a residual 
flavour-paring force, the so-called 't Hooft interaction,  
which leads to good results for meson and baryon mass spectra within 
a nonrelativistic constituent quark model [18]. A covariant BS
 framework for mesons, which also takes the 't Hooft 
interaction into account has been investigated recently by M\"unz [19].
In the treatment of M\"unz the BS equation is formulated 
in the instantaneous approximation (Salpeter equation), and a remarkable 
success is achieved by the description of the meson spectrum and 
dynamical observables. To our knowledge, till now there is no relativistic 
treatment of nucleon form factors which includes the 't Hooft 
interaction in the baryon dynamics. In this sense the investigations
presented here seems quite worthwhile.

The case of baryons within the BS framework 
is technically and also conceptually much more involved than mesons.
Nevertheless, we show in the next section that under some assumptions  
one can arrive at a reduced Salpeter equation for baryons, which has 
the same form as the eigenvalue equation of the mass operator obtained 
in the RHD.  Therefore, we are able to identify the mass operator of 
the RHD with the Salpeter Hamiltonian, and the 't Hooft interaction used 
in the Salpeter framework can be extended to investigate baryons within the RHD.   

This paper is organized as follows: in Sec. II the eigenvalue equation of 
the mass operator in the RHD [14] is shown to
be connected to a reduced Salpeter equation, which is derived under 
the assumptions of free quark propagators and collective instantaneous
interaction kernel. In this sense, we identify the mass operator with
the Salpeter Hamiltonian with a linearly rising confining potential 
and a residual 't Hooft-type interaction. 

The nucleon form factors are
computed in Sec. III with the same formalism 
described by Capstick and Keister [12] to investigate the baryon 
electromagnetic current in the impulse approximation. 
Our calculations are performed in the spacelike region up 
to momentum transfer $-q^2=3$ GeV$^2$ and with light-front wave funtions expanded 
up to the sixth harmonic-oscillator quanta. 
We obtain a remarkable description for the proton electric form 
factor $G_E^p$ in this region. The other nucleon form factors are 
shown to be in reasonable agreement with the 
experimental data. We also examine the role of the 't Hooft interaction 
in the model. The configuration mixing generated by this interaction 
is  shown to be particularly important by the description of the neutron 
electric form factor $G_E^n$.  

A summary is given in Sec. IV, where we also point 
out possible improvements and other applications of the model.

\section{The Model}

\subsection{The eigenvalue equation of the mass operator}
In the relativistic Hamiltonian formulation from Keister and Polyzou [14] 
the dynamics is solved by diagonalizing the eigenvalue equation of the mass 
operator, which reads in the case of baryons like
\begin{eqnarray}
M\left|M j \light{P} \mu\right> = \left(M_0 + V\right)\left|M j \light{P} \mu\right>,
\label{MassEq} 
\end{eqnarray} 
where $M_0=\sum_{i=1}^3\omega_i=\sum_{i=1}^3\left(\sqrt{{\bf k}_i^2+m_i^2}\right)$
is the kinetic energy operator, ${\bf k}_i$ and $m_i$ are the constituent quark 
momentum and mass respectively, $V$ is the interaction that fulfills the 
conditions of the Bakamijan-Thomas construction [20], and 
$\left| M j \light{P} \mu \right>$
is the eigenstate  of the mass $M$, total spin $j$, total 
momentum $\light{P}=(P^+,{\bf P}_\perp)$ and total spin projection $\mu$ operators.
Throughout this paper the same notation as in  Ref. [14], with 
light-front coordinates \linebreak 
$p^- = p^3 - p^0$,  ${\bf p}_\perp = (p^1,p^2)$ and $p^+ = p^3 + p^0$,
is used.

We denote here the irreducible states of the free three-particle Hilbert space by
$\left|[a]({\bf k}_\xi,{\bf k}_\eta)j\light{P} \mu\right>$, where $[a]$ is the set of 
internal spins and angular momenta with a given coupling scheme, 
and $({\bf k}_\xi,{\bf k}_\eta)$ are the relative momenta. This space is
suitable to represent the above eigenvalue equation, because its elements are
also eigenstates of $P$, $j$ and $\mu$.

The Bakamjian-Thomas construction yields, in the irreducible basis, 
 the following condition for the interaction $V$ [14]
\begin{eqnarray}
\left<[a'] ({\bf k}'_\xi,{\bf k}'_\eta)j' \light{P}' \mu'\right| 
V \left|[a]({\bf k}_\xi,{\bf k}_\eta)j
\light{P} \mu\right> &=& \delta_{j'j} \delta_{\mu' \mu} (2\pi)^3
\delta(\light{P}'-\light{P}) \nonumber \\
&& \;\;\;\times \left<[a']({\bf k}^{\prime}_\xi,{\bf k}^{\prime}_\eta)\left\| V^j 
\right\|[a]({\bf k}_\xi,{\bf k}_\eta) \right>.
\end{eqnarray}

Therefore, the eigenvalue equation for the mass operator
(\ref{MassEq}) can be put in the following form
\begin{eqnarray}
\left( \left(\omega_1 +\omega_2 +\omega_3 + V^j \right) \Psi^j_M \right) 
([a]({\bf k}_\xi,{\bf k}_\eta)) = M \Psi^j_M([a]({\bf k}_\xi,{\bf k}_\eta)),
\label{evemass}
\end{eqnarray}
where the wave funtion $\Psi^j_M$ is defined through 
\begin{eqnarray}
\left<[a]({\bf k}_\xi,{\bf k}_\eta) j' \light{P}' \mu'\right. \left|M j \light{P}
  \mu\right>=
  \delta_{j'j}\delta_{\mu'\mu}(2\pi)^3\delta(\light{P}'-\light{P}) 
\Psi^j_M([a]({\bf k}_\xi,{\bf k}_\eta)).
\label{WF}
\end{eqnarray}

This formulation of the RHD  introduces some restrictions on the 
interaction, but by no means determines the explicit form of
$V$. \pagebreak As we are
particularly interested in the study of instanton effects, we prefer
to identify our interactions with the Salpeter Hamiltonian, because  
the good results achieved in the case of mesons [19].

\subsection{The reduced Salpeter equation}
A three-quark bound state with total momentum $P$ is described in 
the relativistic quantum field theory by the BS amplitudes [21,22]
\begin{eqnarray}
\label{BSamplitude}
\chi_{P\alpha_1\alpha_2\alpha_3}(x_1,x_2,x_3) &=& \left<0 \left|
T\psi_{\alpha_1}(x_1)\psi_{\alpha_2}(x_2)\psi_{\alpha_3}(x_3) 
\right| P\right>, \\
\bar{\chi}_{P\alpha_1\alpha_2\alpha_3}(x_1,x_2,x_3) &=& 
\left<P \left|
T\bar{\psi}_{\alpha_1}(x_1)\bar{\psi}_{\alpha_2}(x_2)
\bar{\psi}_{\alpha_3}(x_3) 
\right| 0\right>,
\end{eqnarray} 
where $\psi_{\alpha_i}(x_i)$ is the field operator for the particle 
$i$ with Dirac, flavor and color indices labeled by $\alpha_i$, and $T$ is 
the time ordering operator.

The amplitude $\chi_{P\alpha_1\alpha_2\alpha_3}$ fulfills in momentum 
space the following BS equation [22]

\begin{eqnarray}
\label{BSE}
\chi_{P}(k_\xi,k_\eta) = S^F_1(k_1)S^F_2(k_2)S^F_3(k_3) 
\int  \frac{d^4k_\xi^{\prime}}{(2\pi)^4} 
\frac{d^4k_\eta^{\prime}}{(2\pi)^4}
(-i)K(P;k_\xi,k_\eta;k'_\xi,k'_\eta) 
\chi_{P}(k'_\xi,k'_\eta),
\end{eqnarray}

\noindent where $S^F_i(k_i)$ is the full fermion propagator of 
particle $i$, and $K$ is the BS interaction kernel. The spinor indices have been 
suppressed, and we have introduced the usual Jacobi momenta to factor out the 
c.m. movement.

The normalization condition is obtained by considering the 
pole contribution of the three-body Green's function associated with the 
BS amplitude (\ref{BSamplitude}) (see, e.g. Ref. [22]), it reads
 
\begin{eqnarray}
\label{NCBSE}
&&\!\!\!\!\!\!\!\!\!\!\!\!\int \frac{d^4k_\xi}{(2\pi)^4}\frac{d^4k_\eta}{(2\pi)^4} 
 \int \frac{d^4k'_\xi}{(2\pi)^4}\frac{d^4k'_\eta}{(2\pi)^4} 
\bar{\chi}_{P}(k_\xi,k_\eta) \nonumber \\
&&\;\times\left.\frac{\partial}{\partial P^0}\left[
I(P;k_\xi,k_\eta;k'_\xi,k'_\eta) 
+ iK(P;k_\xi,k_\eta;k'_\xi,k'_\eta)
\right]\right|_{P^0=\omega_P}
\chi_{P}(k'_\xi,k'_\eta) = 2i\omega_P,
\end{eqnarray}

\noindent with $I=(2\pi)^8\delta(k_\xi-k'_\xi)\delta(k_\eta-k'_\eta)[S^F_1(p_1)]^{-1}[S^F_2(p_2)]^{-1}[S^F_3(p_3)]^{-1}$.

Our reduced Salpeter equation is derived with the following assumptions:
\begin{itemize} 
\item The full fermion propagator $S^F_i(k_i)$ is
taken as the free one with only the component that propagates forward in time  
\pagebreak 
\begin{eqnarray}
S^F_i(k_i) \approx i\left(\frac{\Lambda^+_i({\bf k}_i)}{k_i^0 - \omega_i
    +i\epsilon}\right)\gamma^0,
\label{propaga}
\end{eqnarray}
with effective constituent quark mass $m_i$, standard projection 
operators \linebreak $\Lambda^{(+)}_i ({\bf k}_i)=
(\omega_i + H_i({\bf k}_i))/(2\omega_i)$, and Dirac
Hamiltonian $H_i({\bf k}_i) = \gamma^0(\fett{\gamma}
\cdot{\bf k}_i+m_i)$. We felt encouraged to accept this assumption
because it  not only yields some technical simplifications, 
e.g. in the normalization condition of the Salpeter amplitudes, 
but it also isolates formally the kinetic energy contribution.
Hence, the consideration of other features of the baryon dynamics, 
e.g. the confinement is left to the interaction kernel.  
This assumption leads to the Tamm-Dancoff approximation (TDA) of the 
BS equation, which is expected to be a reasonable approximation because 
the nucleon is not treated as a deeply bound-state. 

\item The interaction kernel $K$ is considered collective and instantaneous 
in the baryon rest frame, i.e.
\begin{eqnarray}
\left. K(P;k_\xi,k_\eta;k'_\xi,k'_\eta)\right|_{P=(M,{\bf 0})} = 
V({\bf k}_\xi,{\bf k}_\eta; {\bf k}'_\xi,{\bf k}'_\eta).
\end{eqnarray}
Since the total momentum $P$ is a conserved quantity, we can 
extend the above approximation to any frame [23] simply assuming that  
\begin{eqnarray}
K(P;k_\xi,k_\eta;k'_\xi,k'_\eta) = 
V(k_{\xi_{\perp P}},k_{\lambda_{\perp P}};k'_{\xi_{\perp P}},k'_{\lambda_{\perp P}}),
\end{eqnarray}
with $k_{i_{\perp P}}=k_i-(Pk_i/P^2)P$.  
\end{itemize}

Some important comments should be made at this point. The
interaction kernel $K$ has contributions of two- and three-body irreducible kernels.
If the two-body kernels are taken instantaneous, the total interaction cannot be put 
in an instantaneous form, because the quark spectator is represented by an
inverse propagator, which depends on the time components $k^0_\xi$ and $k^0_\eta$.    
This is shown diagrammatically in Fig. 1. Therefore, the idea of our 
approximation is to assume that the kernels and their respective spectators
combine to build an effective instantaneous interaction. In this sense
our interaction can be called collective. For instance,
a similar argument has already been used by Mitra and Santhanam [24]. 
Nevertheless, they use two-body instantaneous kernels and suppress 
the non-instantaneity due to the quark spectator by introducing 
appropriate $\delta$-functions for the time-component 
of the relative momenta.

With the above assumptions the BS equation (\ref{BSE}) 
can be integrated out over the time-component of the relative 
momenta. Therefore, we get a reduced Salpeter equation, which can be written 
as an eigenvalue equation like 
\begin{eqnarray}
\label{3PSEeigenvalue}
 ({\cal H}\Phi_M)({\bf k}_\xi,{\bf k}_\eta) &=& M \Phi_M({\bf k}_\xi,{\bf k}_\eta) 
\nonumber \\
& =& (\omega_1 + \omega_2 + \omega_3)\Phi_M({\bf k}_\xi,{\bf k}_\eta) \nonumber \\  
& +& \Lambda^{+++} (\gamma^0 \otimes \gamma^0
\otimes \gamma^0) \int \frac{d^3{\bf k}'_\xi}{(2\pi)^3} \frac{d^3{\bf k}'_\eta}{(2\pi)^3} 
 V({\bf k}_\xi,{\bf k}_\eta;{\bf k}_\xi^{\prime},{\bf k}_\eta^{\prime})
\Phi_M({\bf k}_\xi^{\prime},{\bf k}_\eta^{\prime}),  
\end{eqnarray}
with the Salpeter amplitude
\begin{eqnarray}
\Phi_M({\bf k}_\xi,{\bf k}_\eta) = \left(\int 
  \frac{dk_\xi^{0}}{2\pi}\frac{dk_\eta^{0}}{2\pi} \chi_P
(k_\xi,k_\eta)\right)_{P=(M,{\bf 0})}
\label{SalpAmpl}
\end{eqnarray}
and the tensor product of projection operators
\begin{eqnarray}
\Lambda^{+++} = \Lambda^+_1({\bf k}_1) \otimes \Lambda^+_2({\bf k}_2)
  \otimes \Lambda^+_3({\bf k}_3).  
\end{eqnarray}

From Eq. (\ref{NCBSE}), we obtain the normalization condition of the Salpeter 
amplitudes, which is given by   
\begin{eqnarray}
\int \frac{d^3{\bf k}_\xi}{(2\pi)^3} \frac{d^3{\bf k}_\eta}{(2\pi)^3} \;\mbox{tr}
\left\{\Phi_M^{\dagger}({\bf k}_\xi,{\bf k}_\eta) \Phi_M({\bf k}_\xi,{\bf k}_\eta)
\right\}=2M.
\label{NC}
\end{eqnarray}

We see that the reduced Salpeter equation (\ref{3PSEeigenvalue}) has the 
same form as the eigenvalue equation of the mass operator (\ref{evemass}). 
Moreover, in the present formulation the Salpeter Hamiltonian ${\cal H}$ is 
explicitly positive definite due the presence of the projection operator 
$\Lambda^{+++}$, and the Eq. (\ref{NC}) yields  real eigenvalues.
Therefore, the Salpeter amplitudes remain in a Hilbert space like the 
wave funtions in the RHD, and we can identify the mass operator 
with the Salpeter Hamiltonian ${\cal H}$.

\subsection{The internal baryon dynamics}

Like the nonrelativistic calculation, we still can examine the internal 
baryon dynamics with basically three components, namely the kinetic energy, 
the confinement and the fine-structure interactions.

Unfortunately, one is not able to derive from the QCD Lagrangian the
Salpeter kernel that describes the confinement mechanism. Therefore,
our confinement kernels must be phenomenologically motivated.  
The experimental analysis of the Regge trajectories within the
Chew-Frautchi plot [25], as well as lattice calculations of
QCD [26] support the assumption of a string-like behavior of 
confinement. The two- and three-body string potentials are given
by   
\begin{eqnarray}
V_{\mbox{\scriptsize{conf}}}^{\mbox{\scriptsize{2-body}}} 
&=& a_2 + b_2 \sum_{i<j}|{\bf x}_i-{\bf x}_j|, \\
V_{\mbox{\scriptsize{conf}}}^{\mbox{\scriptsize{3-body}}} 
&=& a_3 + b_3 \min\limits_{{\bf x}_0} 
\left(\sum_{i=1}^3|{\bf x}_i-{\bf x}_0|\right).
\end{eqnarray}
It is well known that the three-body potential
can be well  approximated by the two-body string [27]. Therefore, the
scalar part of the confinement potential is  parameterized in this work
just with a two-body string potential. This local potential yields 
a convolution-type kernel in momentum space. 

There is also no precise candidate for the Dirac structure of the
confinement potential from pure QCD analysis. The phenomenological
analysis of the Salpeter framework for mesons [19] shows that the 
scalar $1 \otimes 1$ spin structure yields  a resonable Regge behavior,
and the timelike vector $\gamma^0\otimes\gamma^0$ 
spin structure reproduces the masses and decays of the low lying mesons.
We make an extension  of this analysis for the case of baryons 
taking a combination of the above structures, which reasonably
describes the Regge behavior in the $\Delta$-sector. 
The confinement potential is parameterized here like 
\begin{eqnarray}
V_{\mbox{\scriptsize{conf}}} ({\bf x}_1,{\bf x}_2, {\bf x}_3)& = &
\sum_{(123)}[a + b(\mid {\bf x}_1 - {\bf x}_2 \mid)] \\
&\times& \frac{1}{2}(1\otimes 1\otimes 1\otimes + \gamma^0
\otimes \gamma^0 \otimes 1).
\label{Vconf}
\end{eqnarray}

Also in the Salpeter framework, the baryon spectrum cannot
 be described with a confining potential alone, and the introduction
 of a residual interaction is necessary. We use here the 't Hooft 
interaction, which is based on instanton effects extracted 
from the nonperturbative soft regime of the QCD [15-17].  
In contrast with the OGE calculations (see e.g. Ref. [28]), 
the 't Hooft interaction is a flavor-dependent pairing 
force. It means that this interaction is able to remove, e.g., 
the $\pi$-$\eta$ and the $N$-$\Delta$ degeneracies, as observed 
in nonrelativistic calculations [18].
Recent calculations with the 't Hooft interaction within a Salpeter
framework for mesons [19] not only yield the correct 
$\pi$, $\eta$ splitting,  but also solve the $n\bar{n}-,$$s\bar{s}$-mixing 
for the $\eta$ meson. Therefore, we felt encouraged to extend such 
consideration for the case of baryons.

The two-body 't Hooft Lagrangian has the following structure
\begin{eqnarray}
\label{lag}
\Delta {\cal L}(2)& = &-\frac{3}{16} \sum_i \sum_{kl} 
\sum_{mn} g_{\mbox{\scriptsize{eff}}}(i)
\epsilon_{ikl} \epsilon_{imn} \nonumber \\
&& \times \left\{ :\bar{q}_k \bar{q}_l\left({\bf 1}\otimes {\bf 1} 
+  \gamma^5\otimes\gamma^5 \right) \left( 2{\sf P}^{\cal C}_{\bar{3}} + {\sf
  P}^{\cal C}_6 \right) q_m q_n :\right\},
\end{eqnarray}
where the sum are over flavor indices and the effective coupling
constant $g_{\mbox{\scriptsize{eff}}}$ depends on the quark flavor, and 
it is taken as a free parameter in our model. The tensor notation 
\begin{eqnarray}
\bar{q} \bar{q} (A \otimes B) q q = \sum_{ij} \sum_{kl} \bar{q}_i
\bar{q}_j A_{ik} B_{jk} q_k q_l
\end{eqnarray}
has been used for Dirac and color indices. The operators ${\sf P}^{\cal
  C}_{6}$ and ${\sf P}^{\cal C}_{\bar{3}}$ are respectively
  color sextet and anti-triplet projectors. They are given by
\begin{eqnarray}
{\sf P}^{\cal C}_{6} &:=&  \frac{2}{3} {\bf
  1}^{\cal C} + \frac{1}{4} \fett{\lambda} \cdot \fett{\lambda}, \\
{\sf P}^{\cal C}_{\bar{3}} &:=& \frac{1}{3} {\bf
  1}^{\cal C} - \frac{1}{4} \fett{\lambda} \cdot \fett{\lambda}, \nonumber
\end{eqnarray}
where $\lambda^a$ $(a = 1,...,8)$ are the $SU(3)$ color matrices. 

The 't Hooft interaction derived using the Wick's theorem and 
the Lagrangian (\ref{lag}) is essentially a two-body interaction [19]. 
It is employed in the three-body Salpeter kernel taking the following 
three-body extension 
\begin{eqnarray}
V_{\mbox{\scriptsize{'t Hooft}}}({\bf x}_1,{\bf x}_2,{\bf x}_3) = \sum_{i<j}
V_{\mbox{\scriptsize{'t Hooft}}}^{ij} ({\bf x}_i-{\bf x}_j),
\label{thooft}
\end{eqnarray}
with
\begin{eqnarray}
V_{\mbox{\scriptsize{'t Hooft}}}^{12} ({\bf x}_1-{\bf x}_2) &=& -4\left(g{\sf P}^{\cal
    F}_{{\cal A}_{12}}(nn) + g' {\sf P}^{\cal F}_{{\cal
    A}_{12}}(ns)\right) \nonumber \\
&& \times \left({\bf 1}\otimes {\bf 1}\otimes {\bf 1} 
+  \gamma^5\otimes\gamma^5\otimes {\bf 1} \right)
 \delta^3({\bf x}_1-{\bf x}_2),
\end{eqnarray}
where $g=\frac{3}{8}g_{\mbox{\scriptsize{eff}}}(n)$ and 
$g'=\frac{3}{8}g_{\mbox{\scriptsize{eff}}}(s)$ are the 
coupling constants. We use $n$ to denote the $u$ and $d$ flavors. 
${\sf P}^{\cal F}_{{\cal A}_{12}}(nn)$  and ${\sf P}^{\cal F}_{{\cal A}_{12}}(ns)$
 are projection operators of flavor states that are anti-symmetric under 
permutation of the first and second quarks. 

The point-like 't Hooft interaction is regularized with a Gaussian
function like 
\begin{eqnarray}
\delta^3({\bf x}_1-{\bf x}_2) \rightarrow \frac{1}{\lambda^3
  \pi^{3/2}} \exp\left(-\frac{|{\bf x}_1-{\bf x}_2|^2}{\lambda^2}\right),
\end{eqnarray}
where the finite range $\lambda$ is taken as a free parameter. Like the 
confinement term, this regularization also yields  a convolution-type kernel 
in momentum space.

\section{Results and discussion}
As we aim to investigate the nucleon form factors, our analysis here 
will concern only the nonstrange sector. Our models contains five parameters, 
namely the nonstrange quark mass $m_{n}$, the constant $a$ and the 
slope $b$ of the confinement potential (\ref{Vconf}), the coupling constant 
$g$ and the effective range $\lambda$ of the 't Hooft 
interaction (\ref{thooft}). We examine two sets of parameters with 
different constituent quark masses, which are shown in Tab. I. 

Equation (\ref{3PSEeigenvalue}) has been solved by diagonalizing the
Hamiltonian in a harmonic-oscillator basis up to 14 quanta and by applying
the Rayleight-Ritz variational principle. The confinement parameters 
$a$ and $b$ have been fixed in such a way as to yield the best Regge behavior 
in the $\Delta$-sector. We have chosen the coupling $g$ and the effective 
range $\lambda$  to reproduce the separation between the ground states
of the nucleon and the $\Delta$. The computed masses in Models 1 and 2 
are shown in Tab. II (in MeV). Experimental data are from Ref. [29]. 
Both models lead to a  reasonable description of the nucleon 
$N\frac{1}{2}^+$ ground state and of the Regge trajectory in the
$\Delta$-sector. The problems with the Roper resonance $N(1440)$
and with the nucleon ground state $N(1535)$ with negative parity are
in fact well known from nonrelativistic constituent quark models 
[13, 18] and the situation here remains basically unchanged. 

The component $I^+$ of the electromagnetic current operator $I^\mu$
contains  all the information necessary to obtain the nucleon 
electromagnetic form factors [14]. This is one of the advantages in 
using light-front coordinates. The nucleon Pauli form factor 
$F_1(Q^2)$ and the Dirac form factor $F_2(Q^2)$ are obtained from
\begin{eqnarray}
\left< Mj \light{P}^{\prime} \mu^{\prime} \right| I^+(0)  \left| Mj \light{P} \mu \right> =
F_1(Q^2) \delta_{\mu^{\prime}\mu} - i(\sigma_y)_{\mu^{\prime}\mu} \sqrt{\frac{Q^2}{4M^2}}F_2(Q^2),
\end{eqnarray}
where $Q^2=-q^2$ and $M$ is the nucleon mass. The experimental results
are usually given in terms of electric and magnetic form factors, $G_E$
and $G_M$ respectively, which  are defined as
\begin{eqnarray}
G_E(Q^2)&=&F_1(Q^2)-\frac{Q^2}{4M^2}F_2(Q^2), \nonumber \\
G_M(Q^2)&=&F_1(Q^2)+F_2(Q^2). \nonumber  \\
\nonumber 
\end{eqnarray}

To compute the electromagnetic current, we  use the same light-front 
framework from Capstick and Keister, which is described in detail 
in Ref. [12]. The current matrix element between initial and final 
nucleon states is expanded in sets of free-particle states as
\begin{eqnarray}
\left< M^{\prime}j;\light{P}^{\prime} \mu^{\prime} \right| 
I^+(0)  \left| M j \light{P}\mu \right> & = &
\int \frac{d \light{p}_1^{\prime}}{(2 \pi)^3} \frac{d \light{p}_2^{\prime}}{(2 \pi)^3}
\frac{d \light{p}_3^{\prime}}{(2 \pi)^3}
\int \frac{d \light{p}_1}{(2 \pi)^3} \frac{d \light{p}_2}{(2 \pi)^3} 
\frac{d \light{p}_2}{(2 \pi)^3} \nonumber \\
&& \times \sum_{\mu_i,\mu_i^{\prime}}
\left< M^{\prime}j;\light{P}^{\prime} \mu^{\prime} \left| \right. 
\light{p}_1^{\prime}\mu_1^{\prime} 
\light{p}_2^{\prime}\mu_2^{\prime} \light{p}_3^{\prime}\mu_3^{\prime} \right> 
\nonumber \\ 
& & \times \left< \light{p}_1^{\prime} \mu_1^{\prime}
 \light{p}_2^{\prime} \mu_2^{\prime} \light{p}_3^{\prime}\mu_3^{\prime} \left| 
I^+(0)  \right| \light{p}_1 \mu_1 \light{p}_2 \mu_2 \light{p}_3 \mu_3 \right> 
\nonumber \\ 
&&\times \left< \light{p}_1 \mu_1\light{p}_2 \mu_2 \light{p}_3 \mu_3 \left| \right. 
M j \light{P}\mu \right>, 
\label{Iplus}
\end{eqnarray}
where the light-front momenta satisfies
$\light{P}=\light{p}_1+\light{p}_2+\light{p}_3$, and they are related
to the set of coordinates $\{{\bf k}_1,{\bf k}_1,{\bf k}_3 \}$ in the
baryon rest frame through the following transformation
\begin{eqnarray}
x_i & = & p_i^+/P^+ , \nonumber \\
{\bf k}_{i \perp} &=& {\bf p}_{i \perp} - x_i{\bf P}_{\perp} , \nonumber \\
k_{i3} & = & \frac{1}{2}\left[x_i M_0 - \frac{m_i^2 
+ {\bf k}_{i\perp}^2}{x_i M_0} \right] , \nonumber
\end{eqnarray}
with $M_0 := \omega_1 + \omega_2 + \omega_3 $ and Jacobi determinant 
\begin{eqnarray}
\left| \frac{\partial(\light{p}_1,\light{p}_2,\light{p}_3)}
{\partial(\light{P},{\bf k}_1,{\bf k}_2)} \right| = 
\frac{p_1^+ p_2^+ p_3^+ M_0}
{\omega_1 \omega_2 \omega_3  P^+} . \nonumber
\end{eqnarray}

The  nucleon state vectors are related to the wave funtions as follows:
\begin{eqnarray}
\left< \light{p}_1 \mu_1 \light{p}_2 \mu_2 \light{p}_3 \mu_4 \left| 
\right. Mj;\light{P}\mu \right> &=& (2\pi)^3 \delta(\light{p}_1+\light{p}_2+\light{p}_3-\light{P}) \left| 
\frac{\partial(\light{p}_1,\light{p}_2,\light{p}_3)}
{\partial(\light{P},{\bf k}_1,{\bf k}_2 )} \right|^{-1/2} \nonumber \\
&& \times \left<{\scriptstyle \frac{1}{2}}\bar{\mu}_1 {\scriptstyle \frac{1}{2}}
\bar{\mu}_2 \right| \left. s_{12}\mu_{12} \right>
\left<s_{12}\mu_{12} {\scriptstyle \frac{1}{2}}\bar{\mu}_3 \right| \left. s \mu_{s} \right> \nonumber \\ 
&& \times \left<l_\xi \mu_{\xi} l_\eta \mu_\eta \right| \left. L \mu_L \right>
\left<L \mu_L s\mu_s \right| \left. j \mu \right>\nonumber \\ 
&& \times Y_{l_\xi \mu_\xi}(\hat{\bf k}_\xi) Y_{l_\eta \mu_\eta}(\hat{\bf k}_\eta)
\Psi_{M}^{(R)j}(|{\bf k}_\xi|,|{\bf k}_\eta|) \nonumber \\ 
&& \times D^{(1/2)\dagger}_{\bar{\mu}_1 \mu_1}[R_{cf}(k_1)]
D^{(1/2)\dagger}_{\bar{\mu}_2 \mu_2}[R_{cf}(k_2)] \nonumber \\
&& \times D^{(1/2)\dagger}_{\bar{\mu}_3 \mu_3}[R_{cf}(k_3)],  
\end{eqnarray}
with the radial part $\Psi_{M}^{(R)j}$ of the wave funtion defined in
(\ref{WF}), and the $SU(2)$ representation $D^{(1/2)}_{\bar{\mu}_i \mu_i}[R_{cf}(k_i)]$ 
for the Melosh rotation [30].
 
In the impulse approximation, the three-quark current operator is written 
as a sum of three single-quark current operators. Therefore, the 
current matrix element  between two nucleon states (\ref{Iplus}) in this
approximation is given by
\begin{eqnarray}
\left< M^{\prime}j;\light{P}^{\prime} \mu^{\prime} \right| 
I^+(0)  \left| M j \light{P}\mu \right> & = &
\int \!\frac{d \light{p}_1}{(2 \pi)^3} \frac{d \light{p}_2}{(2 \pi)^3} 
\frac{d \light{p}_3}{(2 \pi)^3} \int \!\frac{d \light{p}'_1}{(2 \pi)^3} 
\sum_{\mu_i,\mu_1^{\prime}}
\left< M^{\prime}j;\light{P}^{\prime} \mu^{\prime} \left| \right. 
\light{p}_1\!+\!\light{q} \mu_1^{\prime} 
\light{p}_2 \mu_2 \light{p}_3 \mu_3  \right> \nonumber \\ 
& & \times \left< \light{p}_1 \!+\! \light{q} \mu_1^{\prime} \left| 
I^+_1(0)  \right| \light{p}_1 \mu_1 \right> 
\left< \light{p}_1 \mu_1\light{p}_2 \mu_2 \light{p}_3 \mu_3 \left| \right. 
M j \light{P}\mu \right> \nonumber \\
&+& \mbox{analog terms for }\light{p}_2\; \mbox{and} \; \light{p}_3 .
\label{IplusImpuls}
\end{eqnarray}

We prefer to use in this initial investigations, with the 't Hooft 
interaction, the most simple {\it Ansatz} for the single-particle
current, namely to consider only the constituent quark charge operators 
without anomalous constituent quark magnetic moments, i.e. 
\begin{eqnarray}
\left< \light{p}^{\prime} \mu^{\prime} \left| I^+_i(0)  
\right| \light{p} \mu \right> = \delta_{\mu^{\prime}\mu} \hat{e}_i,  
\end{eqnarray}
where $\hat{e}_i$ represents the quark charge operator.

All calculations has been performed up to 6 harmonic-oscillator shells. 
The multidimensional integrals have been calculated with a {\it VEGAS} 
integration routine [31]. In order to examine the 
relevance of the components of the wave funtion to the form factors, 
we calculate first the proton and neutron form factors taking 
different number of components. 
In Fig. 2 the curve (a) represents the proton electric
form factor $G_E^p$ with Model 1, and the curves (b) and (c) represent the same
calculation, but taking into account only  the largest and the two largest 
components of the proton wave funtion respectively. One sees that the calculation with 
only $2$ components almost represent the full result. In the case of 
the neutron the situation is a little different. We show in 
Fig. 3 the neutron electric form factor $G_E^n$
with Model 1. The curve (a) represents full result, and the curves (b),
(c) and (d) represent the results taking into account only the largest,
the two largest and the three largest components of the neutron 
wave funtion respectively.
We observe that just after considering the third component of the
wave funtion the form factor become positive. In fact, the first two
components are S-waves, and particularly in the neutron case the
inclusion of other wave types seems to play an important role.

We present in Fig. 4 and in Fig. 5 
the proton electric form factor $G_E^p$ and magnetic 
form factor $G_M^p$ (in $\mu_N$) respectively. Our calculations 
are compared with experimental data taken from Ref. [32, 33]. The oscillator
parameter $\beta$ has been chosen in such a way as to reproduce the 
proton electric form factor. Its value is about $0.5$ fm in both
models and it stay in the region of the stable solutions.
In order to investigate the effect of the 't Hooft interaction, 
we switched off the strength constant $g$ and calculated again 
the form factors using the Model 1 and the same scale parameter 
$\beta$. In Fig. 4 and Fig. 5 the 
curves (a) and (b) are obtained with Models 1 and 2 respectively, 
and the curves (c) represent the results with Model 1 without 
the 't Hooft interaction. 

The same form factors are presented in Fig. 6  
and in Fig. 7 for the neutron. Experimental 
data are taken from Ref. [1, 3, 34-36].  The curves (a) and (b) are again 
obtained with Models 1 and 2, respectively, and the curves (c) 
represent the results with Model 1 without 't Hooft interaction. 
We note that the result for the neutron electric form factor 
is the most sensitive to the 't Hooft interaction. Without this 
interaction the neutron wave funtion contains almost
just S-waves, and the form factor behaves like in
Fig. 3 (curves (b) and (c)). The 't Hooft interaction 
induces some configuration mixing due to diquark correlations. 
We did not know much about the effects from  this interaction 
on form factors up to now. What we see from the present model is 
a quite interesting result, namely the configuration mixing yields 
the correct sign of the neutron electric form factor $G_E^n$.

\section{Summary and outlook}
A phenomenological relativistic constituent quark model, 
which includes instanton-induced forces has been introduced 
to evaluate the electromagnetic nucleon form factors. 
We have performed our calculation within a light-front RHD in 
the impulse approximation. The internal baryon dynamics have 
been  identified with  the Salpeter Hamiltonian. It is composed of 
a relativistic kinetic energy part, a confinement potential 
and a residual 't Hooft interaction.  With five parameters,
namely the nonstrange quark mass $m_{n}$, the constant $a$ and the 
slope $b$ of the confinement potential, the coupling constant 
$g$ and the effective range $\lambda$ of the 't Hooft 
interaction (Tab. I), we have obtained a very good description of the
proton electric form factor $G_E^p$ up to $Q^2=3$ GeV$^2$.
The other nucleon form factors are reasonably described.

In spite of considering no constituent quark form factor,
the magnetic moments come out only about 10$\%$ too small for the proton and
about 15$\%$ too small for the neutron. A general better description of 
the magnetic form factors can be expected with the consideration 
of constituent quark form factors (see e.g. Ref. [9, 11]).
Calculations beyond the impulse approximation should also improve the 
results. They are in general very involved and rarely discussed 
in the literature.

We have examined the effects of the 't Hooft interaction on the
nucleon form factors. It has been shown that, if this interaction 
is switched off, our results are very similar to the results with 
a symmetric wave. We have suggested, that the diquark correlations 
generated by the 't Hooft interaction represent a very important 
component for the description of the neutron electric form 
factor $G_E^n$.   

It should also be interesting to extended the investigations presented here
 to compute other baryon dynamical observables of great interest, 
like the proton axial-vector form factor, 
the electroexcitation helicity amplitudes 
of the nucleon- and the $\Delta$-resonances, the proton polarizabilities, 
and the form factors in the strange sectors.

\bigskip

{\bf Acknowledgments:} I would like to thank H. R. Petry and B. C. Metsch for
many helpful discussions. This work was partly supported by the {\sl Deutscher
Akademischer Austauschdienst} (DAAD) and by the foundation 
{\sl Coordena\c c\~ao de Aperfei\c coamento de Pessoal de N\'{\i}vel 
Superior} (CAPES). 

\newpage
\begin{center}
REFERENCES
\end{center}

\noindent [1] E. E. W. Bruins {\it et al.}, Phys. Rev. Lett. {\bf 75}, 21 (1995)\\
\noindent [2] R. C. Walker, Phys. Rev. D {\bf 49}, 5671  (1994)\\
\noindent [3] A. Lung {\it et al.}, Phys. Rev. Lett. {\bf 70}, 718 (1993)\\
\noindent [4] See e.g. the CEBAF experiments, 
G. Petratos and J. Gomes,  CEBAF E-93-024 (1993);
D. Day, CEBAF E-93-026 (1993);
C. Perdrisat and V. Punjabi, CEBAF E-93-027 (1993);
R. Madey and B. Anderson, CEBAF E-93-038 (1993);
W. Brooks and M. Vineyard, CEBAF E-94-017 (1994);
W. Korsch and R. Mckeown, CEBAF E-94-021 (1994)\\
\noindent [5] K. Goeke, A. Z. G\'orski, F. Gr\"ummer, Th. Mei\ss ner, H. Reinhardt, and R. W\"unsch, Phys. Lett. B {\bf 256}, 321 (1991)\\
\noindent [6] C. V. Christov, A. Blotz, K. Goeke, P. Pobylitsa, V. Petrov,
M. Wakamatsu, and T. Watabe, Phys. Lett. B {\bf 325}, 467 (1994)\\
\noindent [7] Ch. Christov, A. Z. G\'orski, K. Goeke, and P. V. Pobylitsa, Nucl. Phys. {\bf A592}, 513 (1995)\\
\noindent [8] V. Keiner, Phys. Rev. C {\bf 54}, 3232 (1996) \\
\noindent [9] F. Schlumpf, Mod. Phys. Lett. A {\bf 8}, 2135 (1993);
J. Phys. G {\bf 20}, 237 (1994)\\
\noindent [10] P.-L. Chung and F. Coester, Phys. Rev. D {\bf 44}, 229 (1991) \\
\noindent [11]  F. Cardarelli, E. Pace, G. Salm\`e, and S. Simula, 
Phys. Lett. B {\bf 357}, 267 (1995); Few-Body Syst. Suppl. {\bf 8}, 345 (1995)\\
\noindent [12] S. Capstick and B. Keister, Phys. Rev. D {\bf 51},
  3598, (1995) \\
\noindent [13] S. Capstick and N. Isgur, Phys. Rev. D {\bf 34}, 2809 (1986)\\
\noindent [14] B. D. Keister and W. N. Polyzou, Adv. Nucl. Phys. {\bf  20}, 225 (1991)\\
\noindent [15] G. 't Hooft,  Phys. Rev. D {\bf 14}, 3432 (1976)\\
\noindent [16] M. A. Shifman, A. I. Vainshtein, and V. I. Zakharov, Nucl. Phys. {\bf B163}, 46 (1980)\\
\noindent [17] H. R. Petry, H. Hofest\"adt, S. Merk, K. Bleuler, H. Bohr, 
and K. S. Narain, Phys. Lett. B {\bf 159}, 363 (1985)\\
\newpage
\noindent [18] W. H. Blask, S. Furui, R. Kaiser, B. C. Metsch, and  
M. G. Huber, Z. Phys. A {\bf 337}, 451 (1990)\\
\noindent [19] C. R. M\"unz, J. Resag, B. C. Metsch, and H. R. Petry,
 Nucl. Phys. {\bf A578}, 418 (1994); Phys. Rev. C {\bf 52}, 2110 (1995)\\
\noindent [20] B. Bakamjian and L. H. Thomas,  Phys. Rev. {\bf 92}, 1300 (1953)\\
\noindent [21] E. E. Salpeter and H. A. Bethe, Phys. Rev. {\bf 84}, 1232 (1951)\\
\noindent [22] Y. Tomozawa, J. Math. Phys. {\bf 24}, 369 (1983)\\
\noindent [23] S. J. Wallace and  V. B. Mandelzweig, Nucl. Phys. {\bf A503}, 
673 (1989)\\
\noindent [24] A. N. Mitra and I. Santhanam, Few-Body Syst. {\bf 12}, 41 (1992)\\
\noindent [25] D. Perkins, {\it Introduction to High Energy Physics} 
(Addison-Wesley, Amsterdam, 1987)\\
\noindent [26] G. Schierholz,  {\it  Fundamental Forces}, edited by
  D. Frame and K. J. Peach (SUSPP Publications, Edinburgh, 1985)\\
\noindent [27] W. H. Blask, Ph.D. thesis, University of Bonn, 1990\\
\noindent [28] M. Beyer, U. Bohn, M. G. Huber, B. C. Metsch, and J. Resag,
 Z. Phys. C {\bf 55}, 307 (1992)\\
\noindent [29] Particle Data Group, Eur. Phys. J. C 3, 613 (1998)\\
\noindent [30] H. J. Melosh,  Phys. Rev. D {\bf 9}, 1095 (1974)\\
\noindent [31] G. P. Lepage, J. Comp. Phys. {\bf 27}, 192 (1978)\\
\noindent [32] P. E. Bosted {\it et al.}, Phys. Rev. Lett. {\bf 68}, 3841 (1992)\\
\noindent [33] G. H\"ohler {\it et al.}, Nucl. Phys. {\bf B114}, 505 (1976)\\
\noindent [34] G. G. Simon {\it et al.}, Nucl. Phys. {\bf A364}, 285 (1981)\\
\noindent [35] S. Platchkov {\it et al.}, Nucl. Phys. {\bf A510}, 740 (1990)\\
\noindent [36] E. B. Hughes {\it et al.}, Phys. Rev. B {\bf 139}, 458 (1965)\\

\newpage

\begin{center}
TABLES
\end{center}

\begin{table}[htbp]
  \begin{center}
    \leavevmode
    \begin{tabular}{ccc}
      Parameter & Model 1 & Model 2 \\ \hline \hline
      $m_{n}$   & 220 MeV & 270 MeV \\
      $a$       & $-$640 MeV & $-$512 MeV\\
      $b$       & 688 MeV/fm & 526 MeV/fm \\
      $g$       & 118 MeV fm$^3$& 118 MeV fm$^3$\\
      $\lambda$ & 0.333 fm & 0.333 fm \\ 
    \end{tabular}
    \caption{Set of parameters in Models 1 and 2.}
    \label{tab:paramset}
  \end{center}
\end{table}

\begin{table}[htbp]
  \begin{center}
    \leavevmode
    \begin{tabular}{cccc}
      Baryon & Model 1 & Model 2 & Experiment \\ \hline \hline
      $N(\frac{1}{2}^+,939)$  & 934  & 935 & 938$-$  939 \\
      $N(\frac{1}{2}^+,1440)$ & 1556  & 1569 & 1430 $-$ 1470 \\
      $N(\frac{1}{2}^-,1535)$ &  1403  & 1404 & 1520 $-$ 1555 \\
      $\Delta(\frac{3}{2}^+,1232)$ & 1221  & 1229 & 1229 $-$ 1235  \\
      $\Delta(\frac{7}{2}^+,1950)$ & 1852 & 1835 & 1935 $-$ 1965 \\
      $\Delta(\frac{11}{2}^+,2420)$ & 2259 & 2240 & 2300 $-$ 2500 \\ 
    \end{tabular}
    \caption{Baryon mass spectrum in  Models 1 and 2 (in MeV). Experimental data are from Ref. [29].}
    \label{tab:Spec}
  \end{center}
\end{table}

\newpage
\begin{center}
FIGURE CAPTIONS
\end{center}

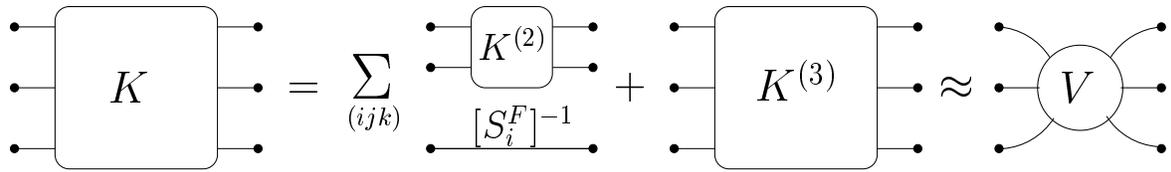
\begin{figure}[htbp]
  \begin{center}
    \leavevmode
    \caption{Diagrammatic representation of the Salpeter kernel $K$, with the 
two- and three-body kernels $K^{(2)}$ and $K^{(3)}$ respectively. 
We approximate $K$ by the collective instantaneous kernel $V$.}
    \label{fig:GEpComp}
  \end{center}
\end{figure}

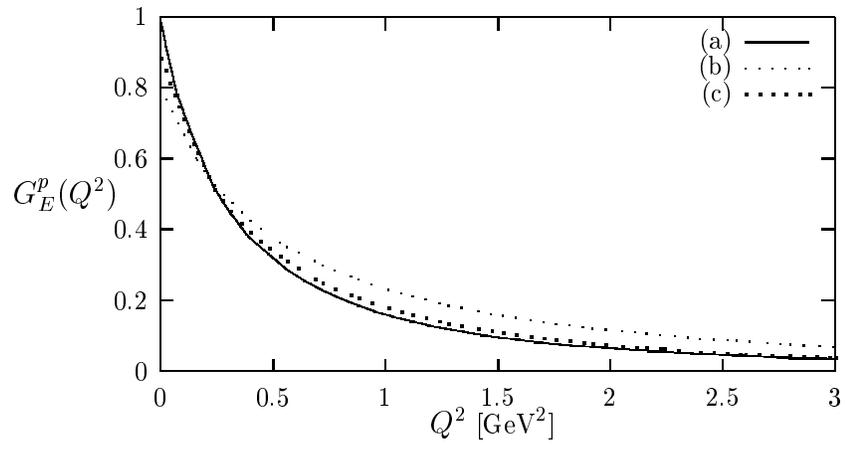
\begin{figure}[htbp]
  \begin{center}
    \leavevmode
    \caption{Proton electric form factor $G_E^p$ with Model 1 (a). The curves
    (b) and (c) represent the results taking into account the largest
    and the two largest components of 
the proton wave function respectively.}
    \label{fig:GEpComp}
  \end{center}
\end{figure}


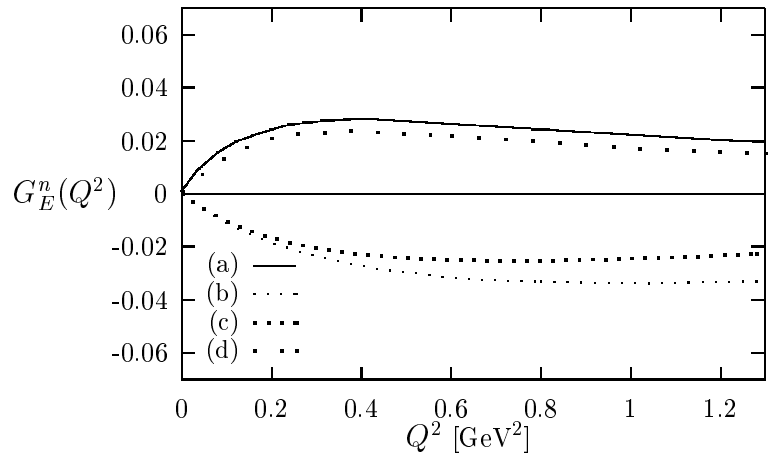
\begin{figure}[htbp]
  \begin{center}
    \leavevmode
    \caption{Neutron electric form factor $G_E^n$ with Model 1 (a). The curves
    (b), (c) and (d) represent the results taking into account only 
the largest, the two largest, and the three largest components of the
neutron wave function respectively.}
    \label{fig:GEnComp}
  \end{center}
\end{figure}


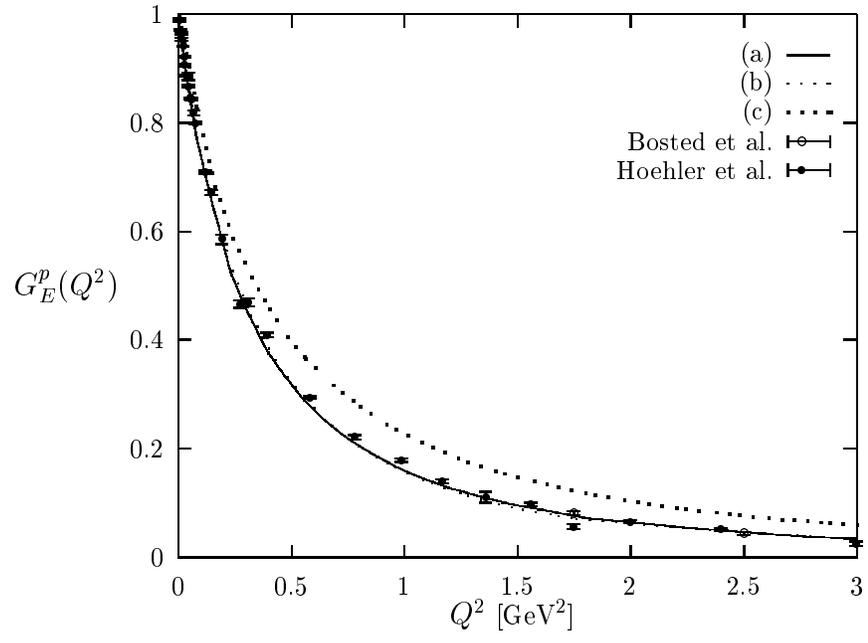
\begin{figure}[htbp]
  \begin{center}
    \leavevmode
    \caption{Proton electric form factor $G_E^p$. The curves (a) and (b) are
    obtained with Models 1 and 2 respectively.  The curve (c)
    represents the results with Model 1 taking into account only the 
    confinement potential, i.e. without 't Hooft interaction. The
    experimental results have been taken from Ref. [32, 33] .}
    \label{fig:GEp}
  \end{center}
\end{figure}


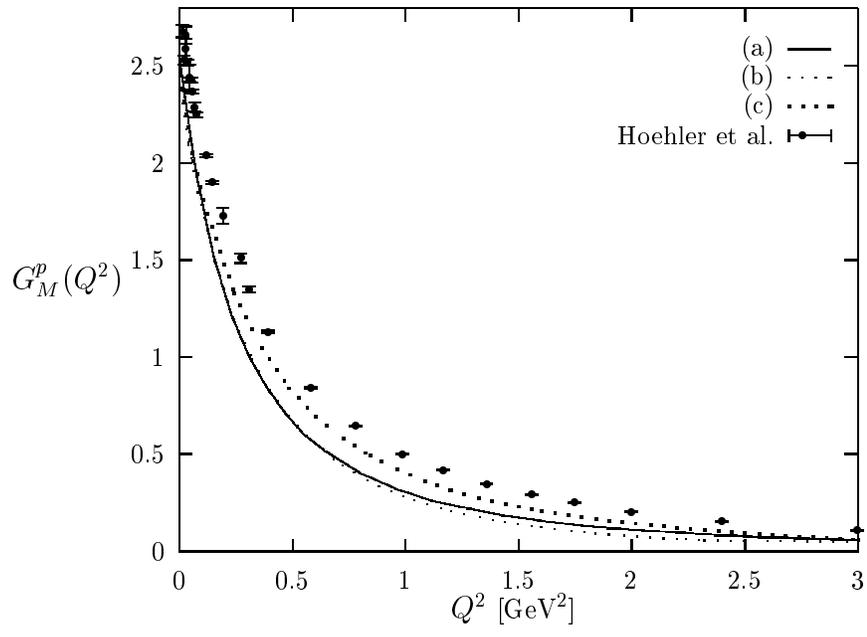
\begin{figure}[htbp]
  \begin{center}
    \leavevmode
    \caption{Proton magnetic form factor $G_M^p$ (in $\mu_N$); key
        as in Fig. 3. Experimental data are from Ref. [33].}
    \label{fig:GMp}
  \end{center}
\end{figure}


\newpage
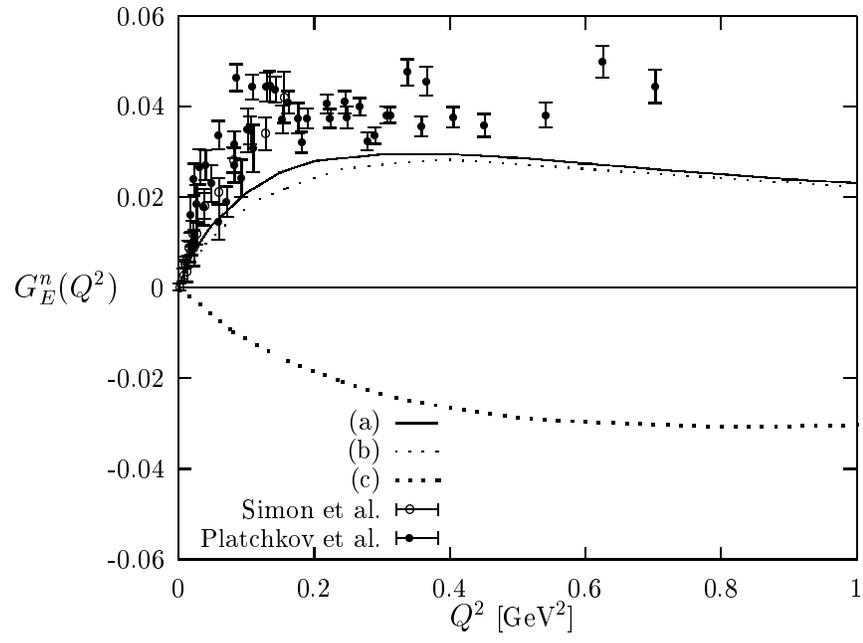
\begin{figure}[htbp]
  \begin{center}
    \leavevmode
    \caption{Neutron electric form factor $G_E^n$; key
        as in Fig. 3. Experimental data are
   from Ref. [34, 35].}
    \label{fig:GEn}
  \end{center}
\end{figure}


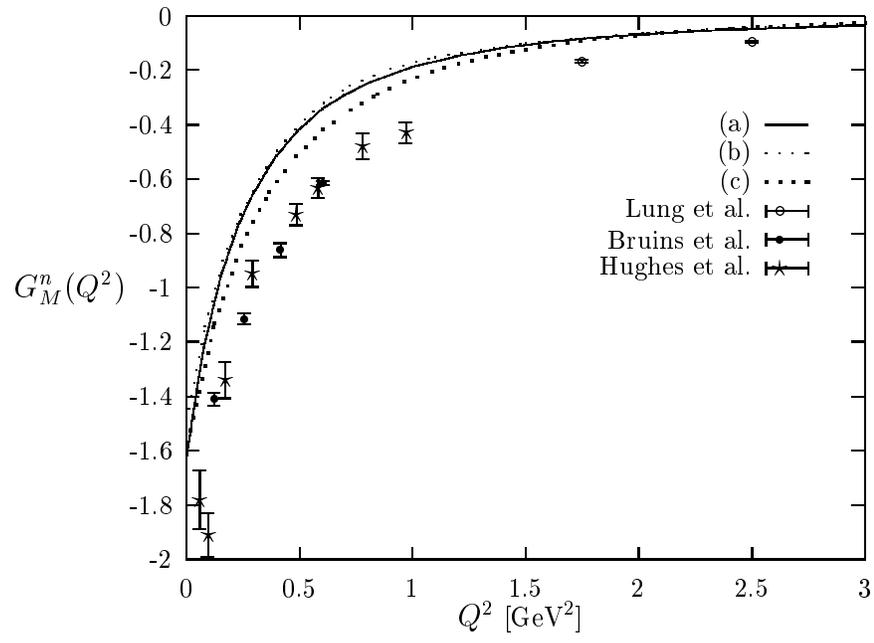
\begin{figure}[htbp]
  \begin{center}
    \leavevmode
    \caption{Neutron magnetic form factor $G_M^n$ (in $\mu_N$); key
        as in Fig. 3. 
  Experimental data are from [1, 3, 36].}
    \label{fig:GMn}
  \end{center}
\end{figure}


\end{document}